# Spectrally-Precoded OFDM for 5G Wideband Operation in Fragmented sub-6GHz Spectrum

R.-A. Pitaval, and B.M. Popović,
*Huawei Technologies Sweden, Stockholm, Sweden*

M. Mohamad, R. Nilsson, and J.J. van de Beek
*Luleå University of Technology, Luleå, Sweden*

## Abstract

We consider spectrally-precoded OFDM waveforms for 5G wideband transmission in sub-6GHz band. In this densely packed spectrum, a low out-of-band (OOB) waveform is a critical 5G component to achieve the promised high spectral efficiency. By precoding data symbols before OFDM modulation, it is possible to achieve extremely low out-of-band emission with very sharp spectrum transition enabling an efficient and flexible usage of frequency resources. Spectrally-precoded OFDM shows promising results for reaching 5G targets in high-data rate enhanced mobile broadband and ultra-reliable low-latency communications use cases. Spectral precoding is particularly efficient for wideband transmission enabling short-time transmission, which will often require flexible fragmented spectrum usage.

## Introduction

Initial commercialization of 5G mobile communications is targeted for 2020. In order to achieve this objective the International Telecommunication Union (ITU) started the IMT-2020 program a few years ago to further develop the International Mobile Telecommunications (IMT) requirements. In early recommendations, 5G is envisaged to support much more data-hungry services than those supported by today's 4G LTE technology. In LTE, allocated bandwidths are limited to 20MHz transmission with an option to aggregate a small number of these. To meet expectations 5G will be required to use wider channels than LTE, possibly up to 1GHz.

As it is difficult to find new frequency bands below 6 GHz a large amount of highly-needed new 5G spectrum is prospected to be found in new millimeter wave bands, *i.e.*, in the range 6GHz - 100GHz, mostly to support short-range radio access. Meanwhile, regulatory actions for these high frequencies will not start before the World Radio Conference (WRC) in 2019, just when 5G early deployments are expected to be launched. Inevitably, these early 5G deployments will therfore be operating in *sub-6GHz* bands and it is widely believed that these lower frequency channels will remain an essential component of 5G even in a longer perspective because their favorable radio propagation characteristics allow ubiquitous and reliable connections. In fact, not only does wideband transmission allow enhanced mobile broadband usage, it is also a key-enabler for ultra-reliable and low-latency communications since lower latency can be achieved by reducing the time transmission and compensated by more frequency resources. As maintaining ultra-reliability may be more challenging in mmWave channels, susceptible to high pathloss and blockage, supporting this application in sub-6GHz band is also crucial.

In the recent WRC of 2015 several new sub-6GHz bands were identified for pre-IMT-2020 [1]. The total spectrum allocated to IMT consists of 15 bands with bandwidths ranging from 20 MHz up to 315 MHz, for an aggregate volume of 1880MHz. Regional variations of the allocation cause the bandwidth and total spectrum to differ across different countries. The allocated spectrum range hosts already many incumbent users and services such as fixed satellite service (FSS), broadcasting satellite service (BSS), radio astronomy, radar system and others, which are taken into account by national regulators. It can be anticipated that 5G Radio Access Technology (RAT) will primarily be specified and deployed in these sub-6GHz bands newly allocated to IMT as they have not been licensed to LTE or UMTS. Deployement to other sub-6GHz IMT bands will subsequently be done by migration from UMTS/LTE.  To this respect,  a global agreement was obtained in WRC-15 to use the spectrum of 694–790 MHz, the L-band (1427–1518 MHz) and in part of the C-band (3400–3600 MHz) for mobile broadband. This results in 387 MHz of globally new spectrum. Additional regional agreements over 618MHz spectrum have also been obtained, including the lower UHF spectrum (470-698 MHz) in Americas while this would remain exclusively allocated to TV broadcasting in ITU Region 1 (Europe, Africa and Middle East) at least until a review in 2023; the extended C-band, 3300-3400 MHz and 3600-3700 MHz in some countries, and the 4800-4990 MHz band for a few countries in Americas and Asia. Finally, early deployment of 5G may also start in unlicensed bands, *e.g* in the 5GHz band.

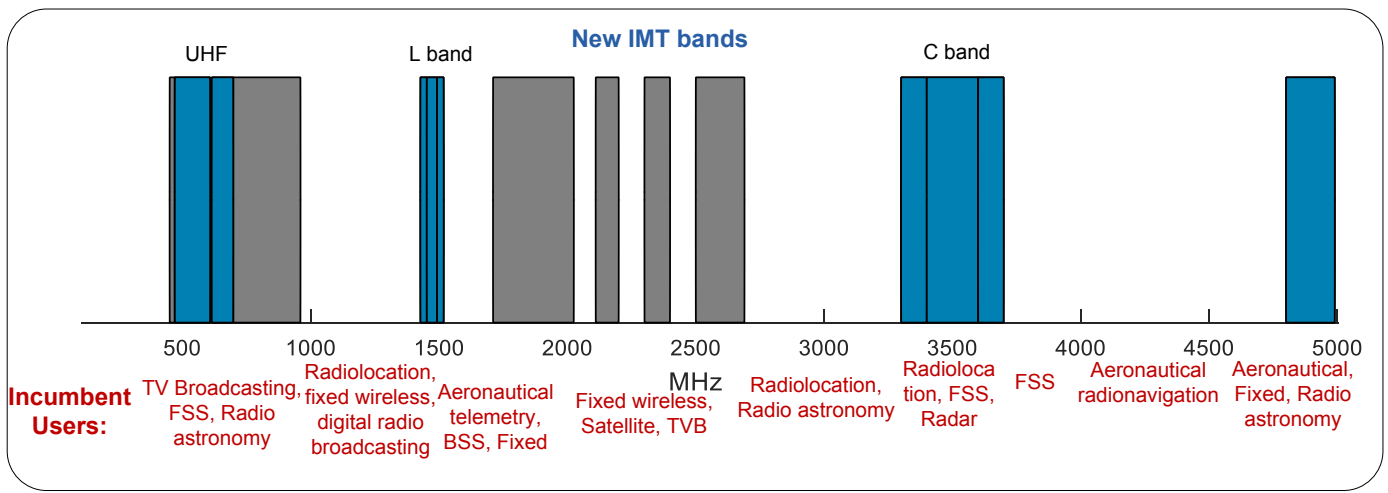

Figure 1. The new bands for IMT/IMT-2020 allocated in November 2015 at WRC.

What are then the 5G challenges in these bands? First, due to the densely packed spectrum in these sub-6GHz bands, high data-rate transmissions require better frequency-localized waveforms than those used in state-of-the-art Orthogonal Frequency Division Multiplexing (OFDM)  transmission in order to support improved spectral efficiency. In LTE, ten percent of a transmission band is lost as a guard band needed to accommodate the system's inherently high out-of-band (OOB) power emissions. Additionally, low-latency application will require the support of larger subcarrier spacing than in LTE and thus also the need for more bandwidth. In brief, 5G will need to be able to efficiently use the available *fragmented* spectrums below 6GHz. It should support wider and more flexible carrier bandwidths than currently defined in LTE to better match the real-world allocation that operators have access to. Spectral masks that specify the maximum levels of OOB emission will likely become increasingly strict with tighter packing of multiple subbands to expand to wideband scenarios through unlicensed and shared-band transmissions. Non-contiguous spectrum will become more predominant as operators, depending on spectrum regulations, will need to combine more bands.

In this paper we present spectrally-precoded OFDM waveforms as a viable means for 5G to address these challenges with spectral efficiency in fragmented sub-6GHz 5G spectrum bands. In subsequent sections we present a brief background to recently emerged variations on the classical OFDM waveform. We conceptually describe OFDM spectral precoding and show that spectrally-precoded OFDM can be used in MIMO communications in a transparent manner by describing and evaluating a novel MIMO-OFDM transmission scheme that incorporates both a spectral and spatial precoder in the transmitter. Spectral precoding may specifically be beneficial for low-latency services and fragmented spectrum usage, and we provide a corresponding comparison with conventional spectrum-shaping methods as low-pass filtering. We conclude this paper with our latest laboratory work and proof-of-concept prototyping results showing the robustness of spectral precoding method against practical-system impairments.

# OFDM-based Waveforms

OFDM is the transmission technology of choice in many of today's communications systems. One disadvantage of OFDM is its relatively large OOB power emission. Out-of-band emissions reduce the overall system performance as they either cause interference in neighboring frequency bands or require maintenance of substantial guardbands to limit this interference.

### *Conventional OFDM*
In 3G/UMTS and 4G/LTE networks, typically, out-of-band emissions are reduced by a filtering or windowing operation at the transmitter. While the 3G standard [2] specifies a root-raised-cosine (RRC) transmit filter with roll-off factor of 0.22, the LTE standard [3] does not explicitly specify a transmit filter. Instead, just a number of signal properties including in-band measures as error-vector magnitude (EVM), and out-of-band performance such as adjacent channel leakage ratio (ACLR), spectrum emission mask (SEM) are specified and the actual signal design is left as a vendor-proprietary choice. While a receiver can thus not rely on a well-defined signal shape, this approach offers a number of transmitter optimization possibilities for the manufacturer.

A low-pass (LP) filtering or windowing operation, typical for LTE transmitters, not only contributes to the overall transmitter EVM, but also introduces intersymbol and intercarrier interference (ISI and ICI) at the receiver [4]. This interference in general comes with a loss in link-SNR, unpredictable and uncontrollable at a receiver. This is more prominent if the channels delay spread is long – with respect to the length of the cyclic prefix. Another drawback of a transmitter low-pass filter is its lack of flexibility to adapt to a variation of transmit scenarios such as, notably, emerging use of non-contiguous spectrum bands.

Fig. 2(a) shows the transmitter chain of low-pass filtered OFDM and two examples of the signal spectra. The spectra correspond to LTE's numerology for the 10MHz bandwidth mode where 600 subcarriers are spaced every 15kHz, thus corresponding to an operational bandwidth of 9MHz along with a 1 MHz guardband. We consider two different types of filter designed to have similar out-of-band suppression and impulse response length. We consider a RRC filter which is a very commonly-used finite impulse response (FIR) filter, with a roll-off factor of 0.22 as required in UMTS, and a filter length tuned to satisfy very stringent OOB reduction. The achieved OOB reduction at the edge of the guard band (at +/- 5MHz, *i.e.*, 0.5MHz away from the outermost subcarriers) is 8dB compared to OFDM. Additionally, we considered an $8^{th}$-order Chebyshev type II filter with similar OOB reduction as such filters were reported to outperform other types of filters [4], generating the least ISI energy under identical complexity constraints. The Chebyshev type-II has an infinite impulse response (IIR), which requires a feedback loop for implementation. While both filters have similar impulse delay, the energy of the Chebyshev filter is much more concentrated in the first instant of the response compared to RRC.

### *Variants of OFDM*
Partly as a results of the emerging interest into flexible and dynamic spectrum usage during the last decade, researchers again have come to challenge the preeminence of OFDM for future cellular systems. For example in non-contiguous spectrum utilization, filters would need to be dynamically designed for each spectrum fragment. Therefore, filterbank multicarrier (FBMC) has been proposed as a alternative for OFDM and has recently been considered as one of the main 5G competing candidate waveform [5]. While FBMC offers very flexible and good time-frequency localization its implementation complexity in a MIMO-system has been identified as a potentially limiting factor. Conventionally low-pass filtered OFDM and FBMC can be viewed as two extreme solutions: in the former a filter is applied to the entire frequency band while in the latter, FBMC, filtering is carried out on a per-subcarrier basis. As a compromise other OFDM variations have been introduced. For example, Universal Filtered Multi-Carrier (UFMC) and f-OFDM [6] applies filtering per sub-band instead of per subcarrier to allow asynchronous transmissions.

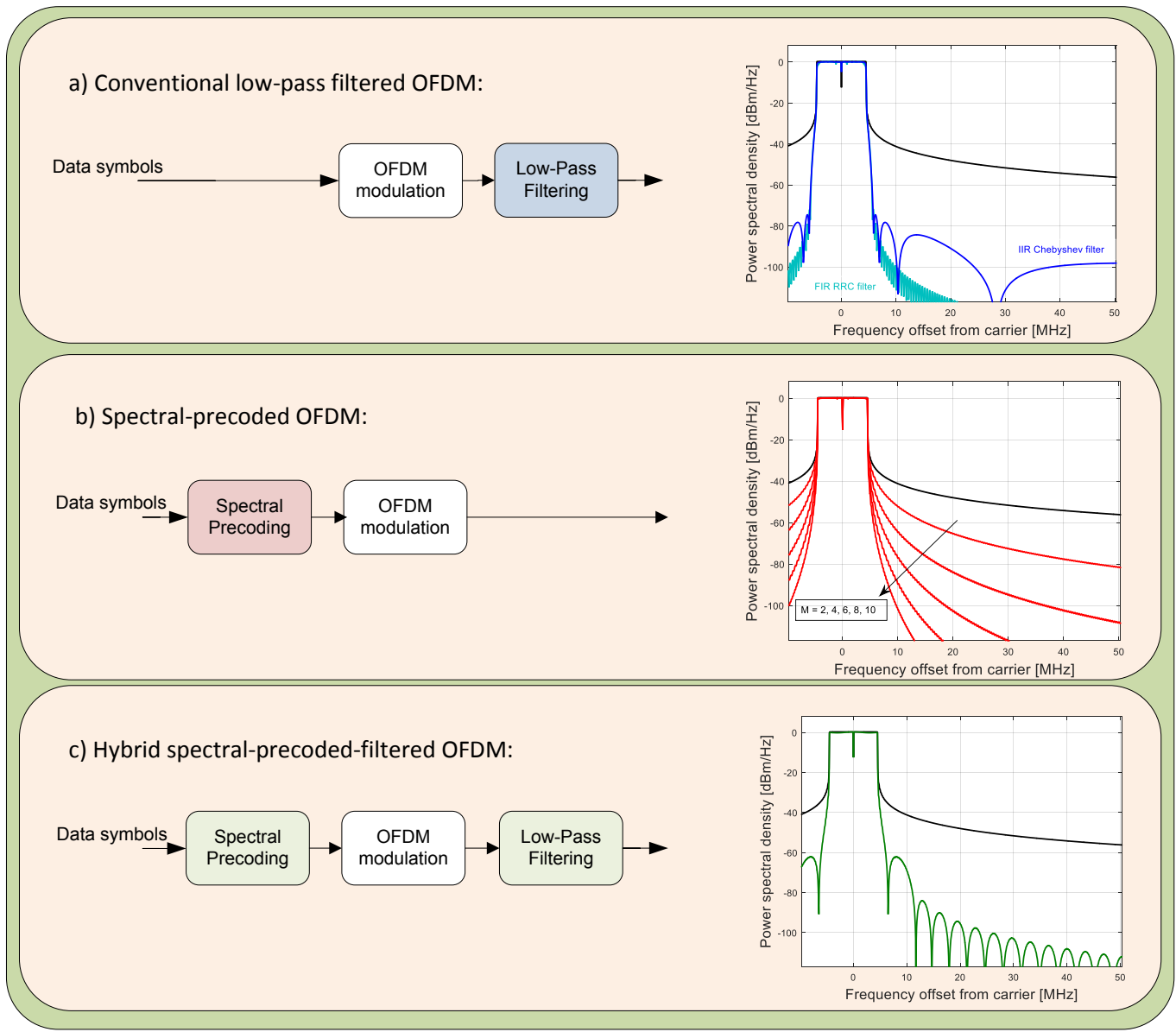

**Figure 2. Different transmitter variations for low OOB emission in OFDM: a) A conventional way to suppress out-of-band emission in OFDM with a spectrum-shaping low-pass filter; b) an alternative way with a spectral precoder; and c) and hybrid version of the two alternatives in a) and b).**

*Spectrally-Precoded OFDM*

Spectral precoding [7] [8] [9] is a transmitter operation where data symbols are modified *prior* to the Fast Fourier Transform (FFT) OFDM modulator in order to relax the requirements, or completely avoid a traditional post-FFT spectrum-shaping low-pass filter, or transmit windowing. While we will describe the concept and implementation in the next section Fig. 2(b) shows the conceptual transmitter chain of a spectrally-precoded OFDM system along with an example of the resulting signal's spectrum. This spectrum is achieved with a precoder which orthogonally projects the 600 data symbols onto a 598-, 596-, 594-, 592-, and 590-dimensional subspace. These subspaces have been selected to guarantee the smoothness of the OFDM signal to be 0, 1, 2, 3, 4-continuous, respectively [10]. The higher-order continuity of the signal trajectory causes the lowest OOB reduction here at the frequency edge of the guard band to be -18dB compared to OFDM.

*Hybrid Spectrally-Precoded Low-Pass Filtered OFDM*

Spectral precoding and low-pass filtering do not preclude each other and can be combined, leading to a hybrid spectrally-precoded low-pass-filtered OFDM system. This hybrid system would relax filter implementation with smoother cut-off frequency transition while a more-relaxed spectral precoding can be applied to sharpen transition between the passband and the stopband of the low-pass filter. Low-pass filtering would guarantee high-frequency OOB reduction while spectral precoding would be used for additional reduction and flexible spectrum shaping, such as in-band frequency holes in carrier aggregations. Fig. 2 (c) shows the conceptual transmitter chain of this hybrid system where a RRC filter with roll-off factor of 0.22 but with an impulse response 6 times shorter than in Fig. 2a). The spectral precoder projects on 596-dimentional subspaces to guarantee a frequency notching at +/- 6.5MHz. The achieved OOB reduction at the edge of the guard band is -10dB compared to OFDM.

## Spectral Projection Precoding: Concept and Implementation

*Transmitter Operation: Projection Precoding*

The fundamental concept of spectral precoding is to confine the pre-FFT transmission to a specific data subspace $\mathcal{G} \subset \mathbb{C}^K$, where $K$ is the number of subcarriers. The chosen subspace is a-priori determined to guarantee a certain level to of low OOB emission.

In the recent literature, several approaches have been developed to restrict transmission to this subspace. Direct modulation of the data symbols in this subspace is one approach, while other approaches adopt a biased or orthogonal projection into the subspace. In this paper, we consider the latter approach, the orthogonal projection method [11], which we believe provides the best alternative and trade-off in term of implementation complexity and performance. The vector of $K$ data symbols is spectrally-precoded *before* feeding the OFDM modulator as

$$\overline{d} = Gd \tag{1}$$

where $G$ is a $K \times K$ orthogonal projection with rank $K - M$ where $M > 0$ is a design parameter. The precoder $G$ namely projects $d \in \mathbb{C}^K$ onto the subspace $\mathcal{G}$ of dimension $(K - M) < K$, so that $\overline{d}$ always belongs to $\mathcal{G}$. Fortunately, for choices of $M \ll K$ subspaces have been described in the literature for which the OFDM signal's OOB emissions are hugely reduced as shown in Fig 2b). The geometry of this projection precoder is shown on Fig. 3a) with its implementation in Fig. 3b). The precoder $G$ is data-independent and changes only when the spectral requirements of the transmitted signal change.

*Constraint Matrix*

One can construct analytically the subspace $\mathcal{G}$ as the nullspace of an $M \times K$ constraint matrix $A$. The number of constraints $M$ corresponds to the dimension reduction. Mathematically the projection precoder is then given by $G = I - A^H(AA^H)^{-1}A$. Since $T = A^H(AA^H)^{-1}$ can be computed in advance, a low-complexity implementation of the precoder emerges from $\overline{d} = d - TAd$ which requires only $2M$ complex multiplications per subcarrier as illustrated in Fig. 3a).

The constraint matrix $A$ can be devised from different methodology. One example is to constrain the transmitted signal to always satisfy $N$-th order continuity [10], a smoothness criterion that guarantees no high frequency components. This method is generic and configurable independently of the transmission parameters. Another method is to *notch* specific well-chosen frequencies which may in-band or out-band depending of specific system-needs [11].

## Fundamental Limit of Spectral Precoding

As shown in Fig. 2, a small number of constraints in the design the spectral precoder, in the order of $M = 10$, is enough to achieve very low OOB emissions. Theoretically, the transmission is analogical to a multi-input multi-output (MIMO) channel which from the rank-deficiency of the projector has a capacity loss of $\frac{M}{K}\%$ compared to OFDM.

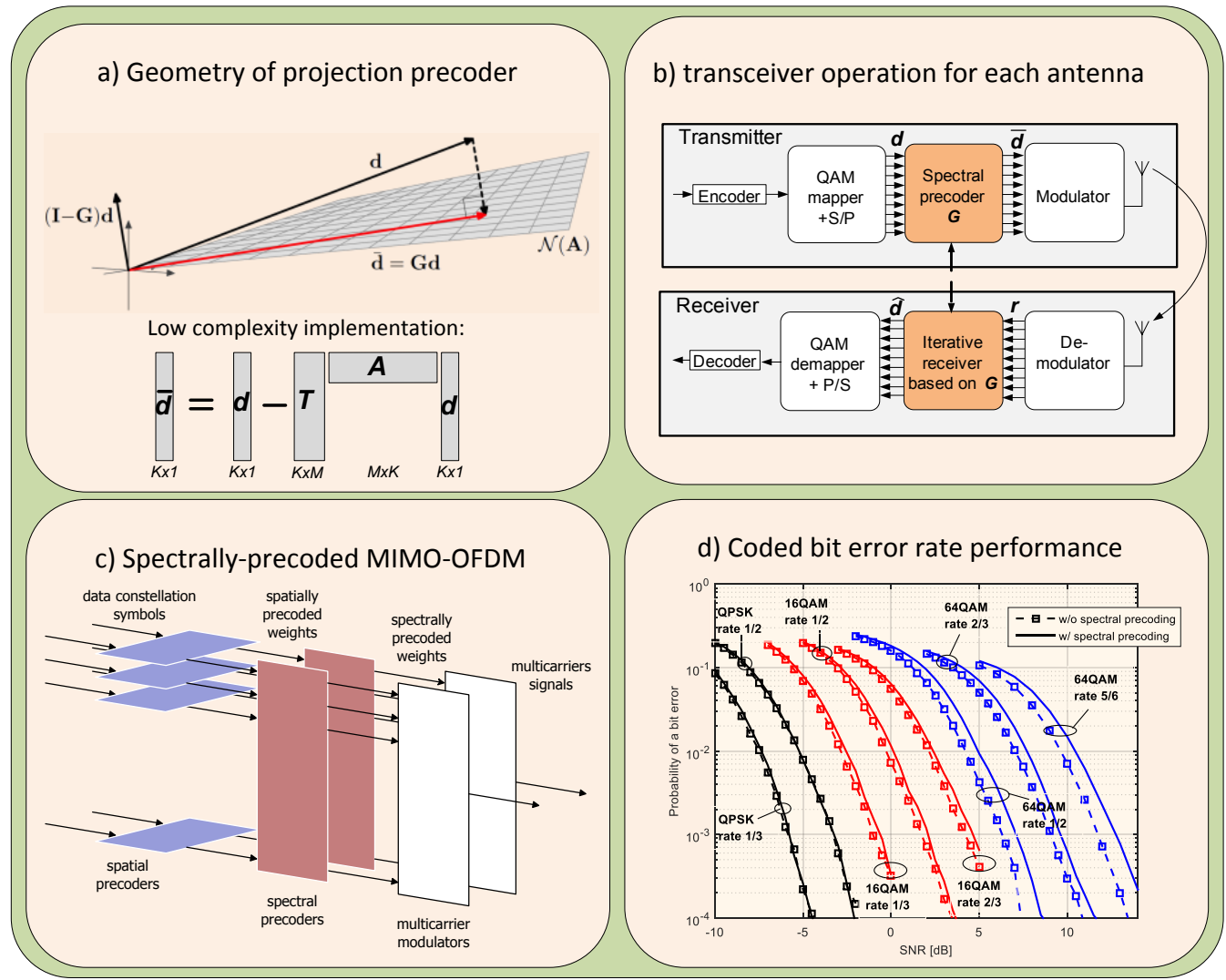

**Figure 3.** The concept of projection precoding and the required transceiver operations. a) The geometry of the precoder. b) The per-antenna transceiver operation. c) A spectrally-precoded MIMO-OFDM transmitter. d) Coded-bit error rate performance with 8Tx antennas.

Therefore, the price to pay to achieve low OOB vanishes as $K \to \infty$, making spectral precoding particularly efficient for wideband transmission with a large number of OFDM subcarriers deployed.

## Receiver Operation: Matched Filtering and Iterative Symbol Estimation

In general, the transmitter projection is a non-invertible operation, that is, the vector $\bar{d} \in \mathcal{G}$ is the image of an infinite number of vectors $d \in \mathbb{C}^K$ that could have been projected onto $\bar{d}$. However, because the transmitted data in $d$ are selected from a discrete constellation, i.e., $d \in \mathbb{C}^K$ where $\mathbb{C}$ is for example a QAM constellation, there is in fact only a finite number of possible $d$ that can be the inverse image of $\bar{d}$. This means that the dimension reduction of the projection precoder may not be an absolute theoretical bottleneck, specifically so for small constellations. However, even if the projection from $\mathbb{C}^K$ to $\mathcal{G}$ would be a one-to-one mapping, inverting $\boldsymbol{G}$ by exhaustive search would be prohibitive complex to implement for large $K$. Nevertheless, the discreteness of the constellation can be exploited in a lower complexity per-subcarrier detection using a non-linear iterative receiver. Here, the first operation on the demodulated received symbols $\boldsymbol{r}$ consists of a matched filter by the spectral precoder $\boldsymbol{G}$ itself . This has the effect of removing by projection the noise term outside of the subspace $\mathcal{G}$. Then, through an iterative symbol estimation a good estimate $\widehat{\boldsymbol{d}}$ of the vector $\boldsymbol{d}$ can be generated. This iterative process shown in Fig. 3b) is improved through the use of soft-symbol estimation based on the noise statistics, and weighted according to the precoder knowledge. Simulations shows that one-to-ten iterations is typically enough.

### Multi-Antenna System

Any new proposal for a future, next generation radio transmission scheme is required to smoothly support multi-antenna transmission. For spectrally-precoded OFDM, such support has not earlier been demonstrated. Yet, the choice of orthogonal projection precoder is particularly relevant for MIMO. The precoder projects the input signals onto a subspace of smaller dimension and hence only slightly distorts the transmitted signal as $\overline{\boldsymbol{d}} = \boldsymbol{d} + \boldsymbol{\epsilon}$. By construction, an orthogonal projection guarantees that this distortion $\boldsymbol{\epsilon}$ has the minimum possible magnitude which is indeed very small. As such spectral precoding can be used transparently with MIMO precoding which is then little affected by this spectral precoder. A novel MIMO-OFDM scheme illustrated in Figure 3c) incorporates this precoding, such that the out-of-band emission from each of its transmit antennas is significantly reduced. In Fig. 3d), resulting coded-bit error rate (BER) performance are shown for this spectrally-precoded MIMO system with the achieved spectrum for $M = 10$ as in Fig. 1b). Several MCS are considered combined with 8 transmit antennas where an optimal spatial precoder is constructed based on perfect channel knowledge. The inevitable price of spectral precoding is paid through a slight reduction of the SNR, here a maximum of one decibel for the highest MCS at $10^{-3}$ BER.

### Spectrally-Precoded SC-FDMA

Single Carrier-Frequency Division Multiple Access (SC-FDMA) system is a channel access method where a transmission is DFT-spread over OFDM subcarriers. A recognized advantage of SC-FDMA over OFDMA is its low peak-to-average-power-ratio (PAPR). As such, SC-FDMA has been chosen as the transmission technology for the LTE uplink. In LTE, consecutive subcarriers are mapped after a DFT-based spreading operation. Spectral precoding can be generalized to SC-FDMA. For this the spectral precoder is either redesigned according to the SC-FDMA waveform or an OFDM spectral-precoder is introduced between the DFT-spreader and the OFDM modulator. Projection precoding barely changes the Peak-to-Average-Power-Ratio (PAPR) of the signal (roughly 0.1dB degradation) and thus the established benefits of SC-FDMA are maintained. Due to the specific single carrier feature of SC-FDMA which corresponds to the multiplexing of timely-shift pulses, low OOB emission imposes to reserve some pulses as guard-pulses to guarantee a smooth transition in time between the SC-FDMA symbols.

## Low-latency Wideband Transmission in Fragmented Spectrum

Among other services 5G will support very low-latency applications that likely will need shorter time-transmission intervals (TTIs) compared to today's LTE. New transmission modes with significantly shorter cyclic prefix will also be investigated to reduce the transmission latency. For example, the length of the cyclic prefix length could be dynamically adapted to the channel time-dispersion rather than being hard-coded in the system's numerology. This small time resource usage can be compensated by more frequency resources in order to maintain sufficiently high data rates. To achieve wider channel transmission, aggregated bands will then more often be needed.

In this section we highlight two particular benefits of spectral precoding for this usage, compared to conventional low-pass filtering. First, spectral precoding performance is independent of the cyclic prefix and transmission time. Second, it can easily shape aggregate multi-band spectrum in order to achieve a wide operational band. These benefits are in sharp contrast with traditional low-pass filters. Interference induced by low-pass filters operation is higher in channels with large delay spreads, and in general, is unpredictable and uncontrollable at the receiver. While for low-latency applications, shorter cyclic prefix lengths may be targeted just long enough t cope with the channel impulse response, the impact of the low-pass filter may then become more pronounced. Furthermore, low-pass filtering lacks the flexibility to adapt to non-contiguous spectrum bands.

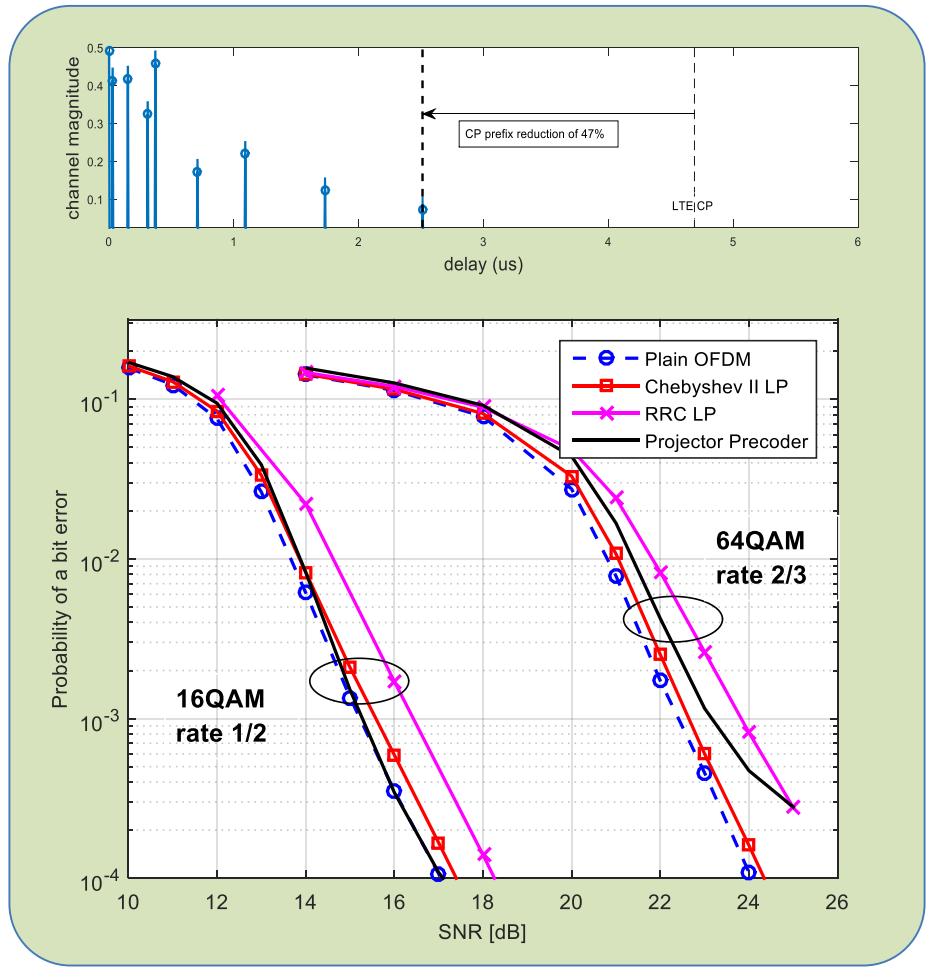

**Figure 4. Comparison of spectral precoding with low-past filtering in in 5G scenarios with low-latency wideband transmission: CP prefix reduction.**

*Cyclic prefix reduction*

Since low-latency is specifically required in vehicular communications, we consider the LTE Extended Vehicular A (EVA) channel. This is consistent with emerging use cases such as mission-critical machine-type communication (MTC), such as self-driving cars or factory environment with moving machines. We consider a 47% reduction from LTE CP, namely the length of the cyclic prefix is set to accommodate the longest delay spread expected on the channel, therefore still eliminating the ISI and ICI.

The impulse response of the EVA channel model and the matching length of the cyclic prefix are shown at the top of Fig. 4. Note that the energy of last tap is very small and most of the energy of the channel is concentrated in the first taps of the impulse response. The resulting performance is shown at the bottom of Fig. 4 for two MSCs, 16QAM rate 1/2 and 64QAM rate 2/3. Without filtering, the cyclic prefix is just long enough to accommodate the delay spread and keep orthogonality between the subcarriers. As subcarriers are obviously affected by frequency selectivity we apply a single-tap zero-forcing equalizer at the receiver, weighing the noise per-subcarrier differently and as a result the BER curves are slightly deteriorated in higher-SNR regions. In the case of low-pass filtering, we assume that the effective channel (physical channel convolved by the impulse response of the LP filter) is known at the receiver and a single-tap equalization with the effective channel is similarly used.

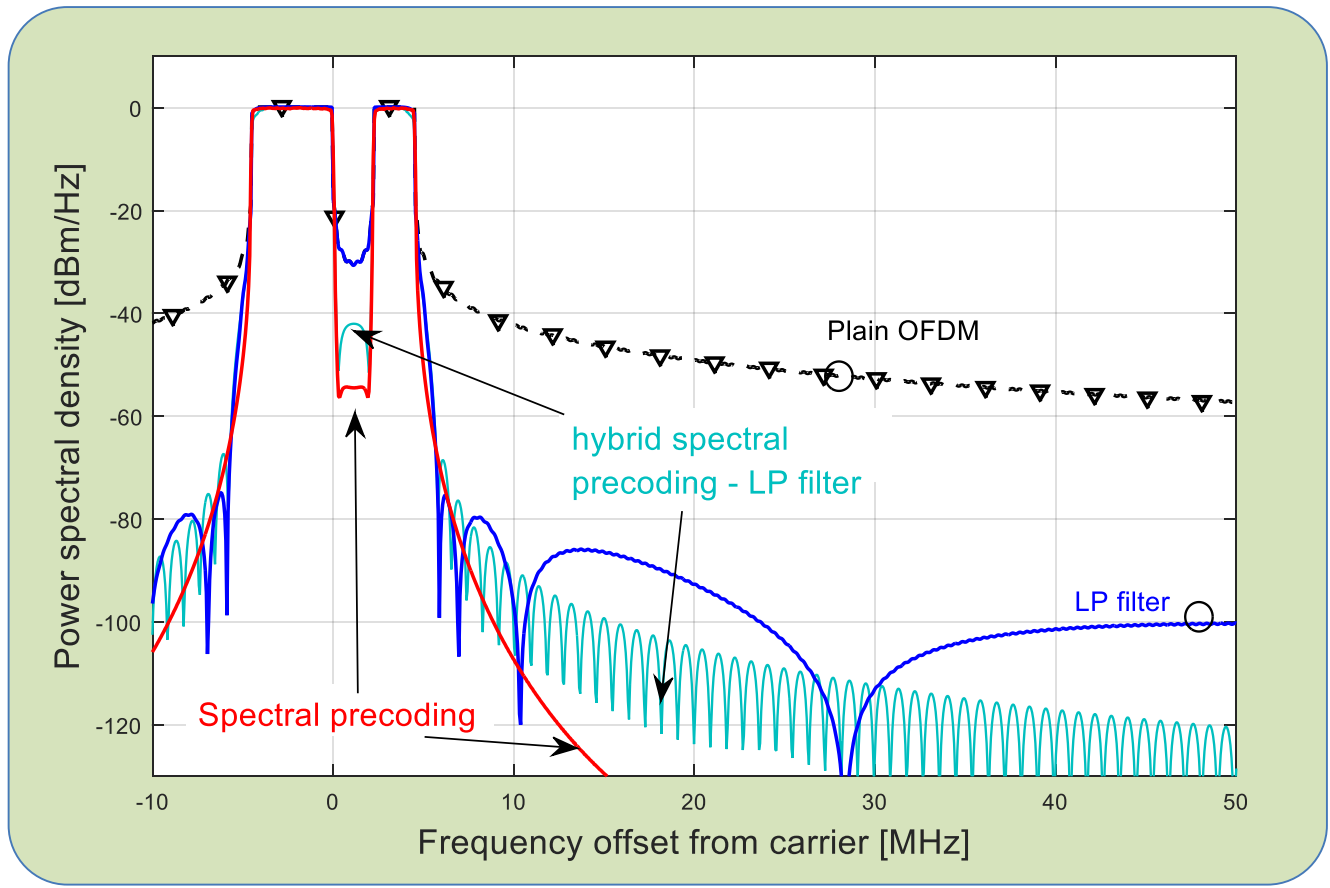

**Figure 5. Comparison of spectral precoding with low-past filtering in in 5G scenarios with low-latency wideband transmission: multiband spectrum shaping for wideband usage.**

As the cyclic prefix is long enough, the performance of plain OFDM is not affected by the duration-reduction of the cyclic prefix. This holds true also for spectral precoding. With this channel, the spectral precoder ($M = 10$) incurs almost no loss in average BER for 16QAM rate 1/2, and for 64QAM rate 2/3 a few decibels loss at $10^{-3}$ compared to OFDM. The performance loss due to the filter depends of the particular choice of filter. We consider two filters with similar out-of-band reduction as shown in Fig. 1. The Chebychev filter provides a better performance with less than half a decibel of SNR loss, while the RRC filter gives a worse degraded performance with 1dB to 1.5dB at $10^{-3}$ BER for both constellation. This is because the impulse response energy of the RRC filter is distributed in a longer time interval than that of the Chebychev filter, although Chebychev has an infinite impulse response. The Chebychev filter performs slightly worse than spectral precoding with 16QAM but slightly better with 64QAM. We conclude that one cannot reduce the cyclic prefix without immediately invoking interference caused by the transmit filter.

*Fragmented spectrum*

Lower latency can also be achieved by reducing the transmission time and increasing transmission bandwidth. Wideband transmissions below 6GHz are prone to encompass fragmented spectrum due spectrum scarcity. One clear advantage of spectral precoding over low-pass filtering is that it offers a flexible spectrum shaping method already available for fragmented spectrum usage. In the case of non-contiguous band allocation, applying a single low-pass filter on the whole OFDM signal does not enable any OOB emission reduction *between* two separated sub-bands. In contrast, spectral precoding allows to control the shape of the power spectrum at will. Spectral precoding can be designed by notching specific frequencies enabling the system to achieve steeply decaying spectral notches.

Fig. 5 illustrates this where a two-band spectrum is created by modulating a single OFDM signal constituted of 450 subcarriers split in two groups of 300 and 150 subcarrriers while 150 subcarriers in a mid-band cannot be used. As expected LP filtering does not reduce OOB emission in the unused subband as this subband is inside the passband of the LP filter. The OOB emission can be reduced by designing spectral precoder that notches specifically well-chosen frequencies as shown in Fig. 5. The spectral notching can be used in a hybrid system with low-pass filtering (here a RRC filter) providing 13dB in the mid-band. Alternatively, the notching can be added in the design of the spectral precoder discussed previously, providing 25dB OOB reduction in the mid-band. In order to achieve OOB reductions in the unused subband with LP filtering only, one would need a filterbank where each filter has a narrower passband. This would cause a longer impulse response of each filter, ultimately causing more inter-symbol interference and thus more BER degradation.

# A proof-of-concept prototype

Finally, we show initial feasibility of the spectral precoding concept for OFDM. A proof-of-concept prototype is being developed at Luleå University of Technology which is illustrated in Fig. 6. The OFDM system implemented on a software-designed radio (SDR) platform of N210 USRP from Ettus research operates at 2.3 GHz, and is based on LTE's 5MHz bandwidth mode. In order to adapt to the bandwidth capabilities of the software radio this mode has been scaled down by a factor of three. The system consists of 300 subcarriers with an intercarrier spacing of 5 kHz (instead of LTEs 15 kHz) and hence occupies an aggregate bandwidth of 1.5 MHz. One quarter of the system bandwidth is ficticiously assumed to be occupied by incumbent services and hence 75 subcarriers are left unmodulated. The spectral precoder of order ($M = 10$) not only suppressed out-of-band emissions but also deepens the notch and hence reduces interference to the incumbent.

The figure shows the spectra of a plain OFDM signal (yellow) compared to the precoded OFDM signal (blue), as measured by a spectrum analyzer. The figure shows a remarkable suppression of the OOB power emission by up to 30 decibels at a frequency 1.2 times the system bandwidth along with an additional 18 decibel suppression in the spectral notch compared to plain OFDM – notably without the use of a low-pass filter.

Compared to the theoretical performance of spectral precoding the efficiency of the precoder has only slightly degraded in the prototype implementation. This degradation appears as a result of the impairments that accompany the hardware implementation. Hardware impairments include power amplifier non-linearities that appear as a drawback of the high dynamic range of the OFDM signal, quantization that approximates the precoded OFDM signal according to the resolution accuracy of the digital-to-analog converter (DAC), clipping of the input precoded signal due to the limited voltage range of the ADC, I-Q imbalances, and to a lesser extent local oscillator impairments that cause a frequency offset between OFDM subcarriers.

Fig. 6 shows that contrary to the theoretical results where the precoded spectrum keeps decaying with frequency, the prototype signal spectrum reaches a typical noise floor that appears, a result of the combined effect of the above impairments. Nevertheless, it is remarkable with all these practical impairments that spectral precoding method is robust enough to provide 15-30 dB OOB reduction compared to plain OFDM.

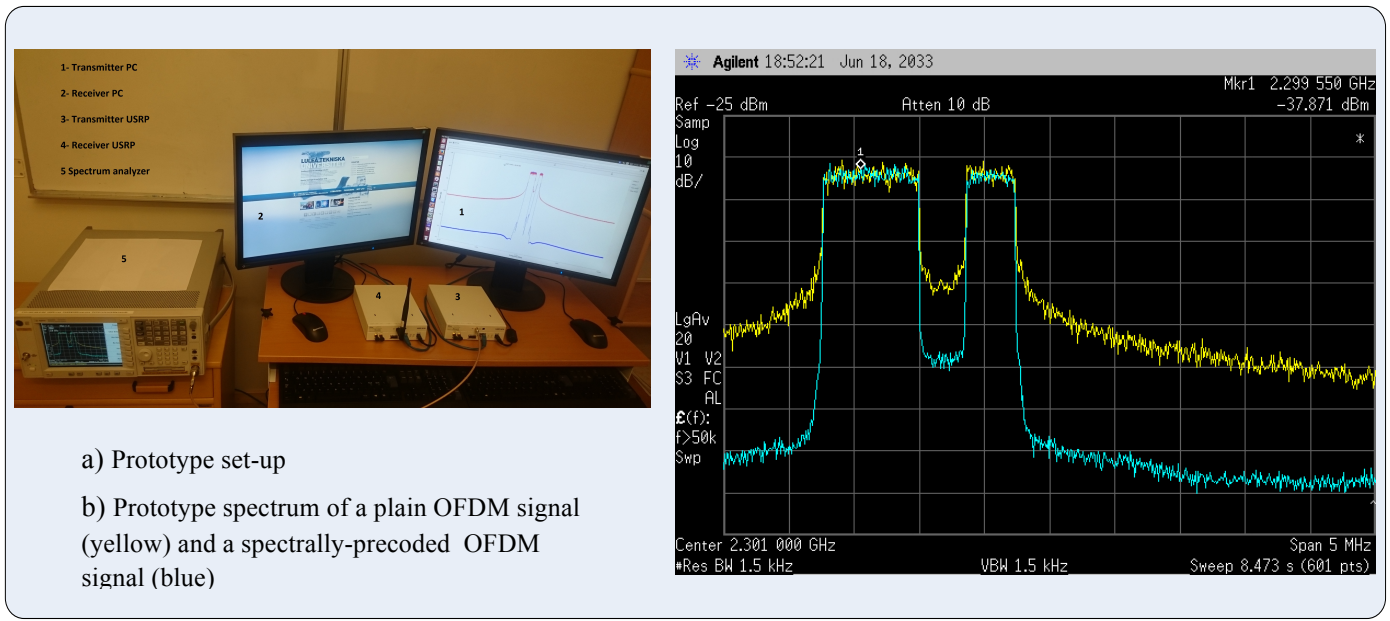

a) Prototype set-up

b) Prototype spectrum of a plain OFDM signal (yellow) and a spectrally-precoded OFDM signal (blue)

**Figure 6. A proof-of-concept prototype of the spectral precoder: theOFDM transmitter deploys a precoder instead of a low-pass filter.**

## Conclusion

Spectrally-precoded 5G wideband operation is an alternative and/or complementary solution to conventional OFDM with a low-pass filter in fragmented sub-6GHz spectrum. It prepares the radio link in a way that has little impact on performance and complexity, while it preserves a large amount of backwards compatibility with LTE specifications. Spectral precoding is an attractive approach for designing 5G waveform with low OOB. This paper provides a conceptual description of spectral precoding including a novel MIMO structure. We show that spectral precoding is able to potentially produce lower SNR loss than some filters in cyclic prefix reduction scenarios, while allowing spectral shaping in heavily fragmented spectrum usage, where a single LP filter cannot provide in-band OOB reduction. Finally, we presented recent proof-of-concept prototyping results that illustrate how well spectral precoding works with imperfect hardware.

## Acknowledgement

The authors with Luleå University of Technology acknowledge the financial support from the Swedish Research Council, grant nr 2014-5977 for parts of the work reported here.